\documentclass[letter]{aa} % for the letters 

\usepackage{graphicx}
\usepackage{txfonts}
\usepackage{xcolor}
\usepackage{mathrsfs}
\usepackage{amssymb,amsmath}
\usepackage{amsmath}    % Advanced maths commands
\usepackage{amssymb}    % Extra maths symbols
\DeclareMathAlphabet{\mathbi}{OT1}{ptm}{bx}{it}
\SetMathAlphabet\mathbi{bold}{OT1}{ptm}{bx}{it}

\usepackage{hyperref}
\hypersetup{
    colorlinks=true,
    citecolor=cyan,
    linkcolor=blue,
    filecolor=magenta,      
    urlcolor=cyan,
}

\def\civ{{C\sc{iv}}\/}
\def\ciii{{C\sc{iii}]}\/}

\begin{document} 

   \title{Accretion disc reverberation mapping in a high-redshift quasar}
   \titlerunning{Accretion disc size in a high-z quasar}
    \author{F. Pozo Nu\~nez \inst{1}\thanks{Corresponding author: francisco.pozon@gmail.com}\href{https://orcid.org/0000-0002-6716-4179}{\includegraphics[scale=0.5]{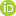}}
          \and
          E. Ba\~nados \inst{2}\href{https://orcid.org/0000-0002-2931-7824}{\includegraphics[scale=0.5]{orcidicon.png}}
          \and
          S. Panda\inst{3}\thanks{Visiting astronomer, MSO/CTIO, NSF NOIRLab}\href{https://orcid.org/0000-0002-5854-7426}{\includegraphics[scale=0.5]{orcidicon.png}}
          \and
          J. Heidt \inst{4}\href{https://orcid.org/0000-0002-0320-1292}{\includegraphics[scale=0.5]{orcidicon.png}}
          }

   \institute{Astroinformatics, Heidelberg Institute for Theoretical Studies, Schloss-Wolfsbrunnenweg 35, 69118 Heidelberg, Germany
   \and
   Max-Planck Institut f\"ur Astronomie, K{\"o}nigstuhl 17 Heidelberg, Germany
   \and
   International Gemini Observatory/NSF NOIRLab, Casilla 603, La Serena, Chile
   \and
   Landessternwarte, Zentrum f\"ur Astronomie der Universit\"at Heidelberg, K{\"o}nigstuhl 12, 69117 Heidelberg, Germany \\  
              }
   \date{Received 7 May 2025 / Accepted 20 July, 2025}

 \abstract{Powered by supermassive black holes at their centers, quasars are among the most luminous objects in the Universe, serving as important probes of cosmic history and galaxy evolution. 
 The size of the accretion disc surrounding the black hole is a critical parameter for understanding quasar physics and their potential use as standard candles in cosmology. 
 However, direct measurements of accretion disc sizes have so far been confined to the Local Universe ($z<0.2$), limiting our understanding of quasars during the peak of cosmic activity. 
 Here, we report the first direct measurement of the accretion disc size in the quasar QSO J0455-4216 at $z=2.66$, when the Universe was only $\sim2$ Gyrs old.
 Medium-band filters mounted on the MPG/ESO 2.2-metre telescope at La Silla Observatory were used to isolate continuum emission regions during a six-month monitoring campaign. 
 The light curves exhibit pronounced variability features and enabled the detection of inter-band time delays from different parts of the disc. 
 We mapped the disc and located its ultraviolet-emitting outermost region at \( 3.02^{+0.33}_{-0.57} \) light-days from the black hole (\( \sim 500 \) AU). 
 Given a supermassive black hole 900 million times the mass of the Sun, these measurements validate accretion disc theory at an unprecedented redshift and pave the way for 
 efficient black hole mass estimates, reducing decades-long spectroscopic reverberation campaigns to just a few years or less.}

   \maketitle

\section{Introduction}

Gas accretion is a fundamental process in the Universe, underpinning many of the phenomena we observe today. From the formation and evolution of cosmic filaments within large-scale structures (\citealt{2019MNRAS.489.2130V}) to the birth of stars and planets (\citealt{2019NatAs...3..749H}) in galaxies, and even the dynamics of compact objects, gas accretion drives cosmic evolution (\citealt{2023ApJ...943...82S,2023MNRAS.525...12S}). At the centres of most galaxies lie supermassive black holes (SMBHs; \citealt{2000ApJ...539L...9F,2023A&A...672A..98M}), which, under certain conditions, can become the most luminous objects in the Universe. These active galactic nuclei (AGNs or quasars) emerge when the SMBH accretes matter at high rates, forming a radiatively efficient and geometrically thin accretion disc (\citealt{2014MNRAS.438..672N}). This phase is critical for understanding the growth of SMBHs throughout the history of the universe since significant mass accumulation can occur during these periods (\citealt{2011ApJ...730....7T}). The Shakura \& Sunyaev (\citealt{1973A&A....24..337S}) accretion disc theory (SS73) is commonly used to describe this process, although other frameworks exist (\citealt{2013LRR....16....1A}). However, accretion discs in high-redshift quasars remain spatially unresolved by current observational technologies, necessitating indirect methods to infer their properties. One such method is reverberation mapping (RM) in its spectroscopic (\citealt{1986ApJ...305..175G,1993PASP..105..247P}) and photometric forms (\citealt{2011A&A...535A..73H,2012A&A...545A..84P}). It  takes advantage of variability, rather than spatial resolution,   significantly advancing our understanding of the size and structure of accretion discs in AGN within the Local Universe. 

Here, we present the results of a dedicated photometric RM campaign focussed on the quasar QSO J0455-4216 ($z=2.66$; \citealt{2018ApJ...865...56L}) conducted using medium-band filters mounted on the MPG/ESO La Silla 2.2 m telescope. The filters were carefully selected to study the variability of the continuum, while minimizing contamination of emission lines from the broad-line region (BLR; \citealt{2017PASP..129i4101P,2019NatAs...3..251C,2019MNRAS.490.3936P}). The high quality of our light curves and the distinct variability features enabled us to examine the temporal structure of the emission and the origins of the observed delays in unprecedented detail. Our analysis reveals that the disc transfer function is capable of reproducing the main features in the observed light curves, with indications that the contributions from the BLR and diffuse continuum regions are limited. This result supports the interpretation that the wavelength-dependent delays observed in QSO J0455-4216 are largely governed by the disc reprocessing geometry, rather than substantial reprocessing in the BLR or diffuse continuum regions.

\section{Observations and data}\label{sec2}

Photometric observations of QSO J0455-4216 were conducted between October 2023 and April 2024 using the Wide Field Imager (WFI) camera mounted on the MPG/ESO La Silla 2.2-meter telescope as part of an ongoing reverberation mapping campaign focussed on high-redshift quasars. 
This luminous and massive quasar ($L_{\rm bol} = 8.25 \times 10^{47} \, \mathrm{erg \, s^{-1}}$ and $M_{\rm BH} = 8.9 \times 10^{8} \, M_{\odot}$) was selected because it was part of one of the longest ($> 10$ years) RM campaigns monitoring the CIV BLR (\citealt{2018ApJ...865...56L}).
The WFI is outfitted with a set of 26 medium-band filters, four of which were specifically selected for these observations based on the redshift of QSO J0455-4216, to capture primarily AGN continuum variations.
These filters have central wavelengths (and FWHM) of 4858 (315), 5315 (255), 6043 (277), and 7705 (257) $\text{\AA}$, enabling efficient observations, while minimizing contamination from emission lines (see Fig.~\ref{fig:AP1}).
The images were reduced using standard procedures for image processing, including bias subtraction, flat-field correction, astrometry, and correction for astrometric distortions. 
The light curves were extracted using aperture photometry relative to six non-variable stars located within a $20'$ radius around QSO J0455-4216.
Figure~\ref{fig:Figure1} shows the normalized light curves of QSO J0455-4216, including a representative reference star in the field. 
The high quality achieved in the reduced images allows us to measure the nuclear flux with a photometric precision of 0.2–0.3 per cent.

In addition to the photometric monitoring, we obtained an optical spectrum of QSO J0455–4216 on 14 February 2024 with the Goodman Spectrograph (\citealt{2004SPIE.5492..331C}), mounted on the 4.1-m Southern Astrophysical Research (SOAR) Telescope. The data reduction included bias subtraction, flat-field correction, cosmic ray removal, sky subtraction, and a wavelength and flux calibration.
We refer to Appendix \ref{sect:data} for more details on the data reduction.
%=========================================================================
\section{Results and discussion}\label{sec3}
%=========================================================================

\subsection{Optical light curves and time delays}

\begin{figure}
    \centering
    \includegraphics[width=\columnwidth]{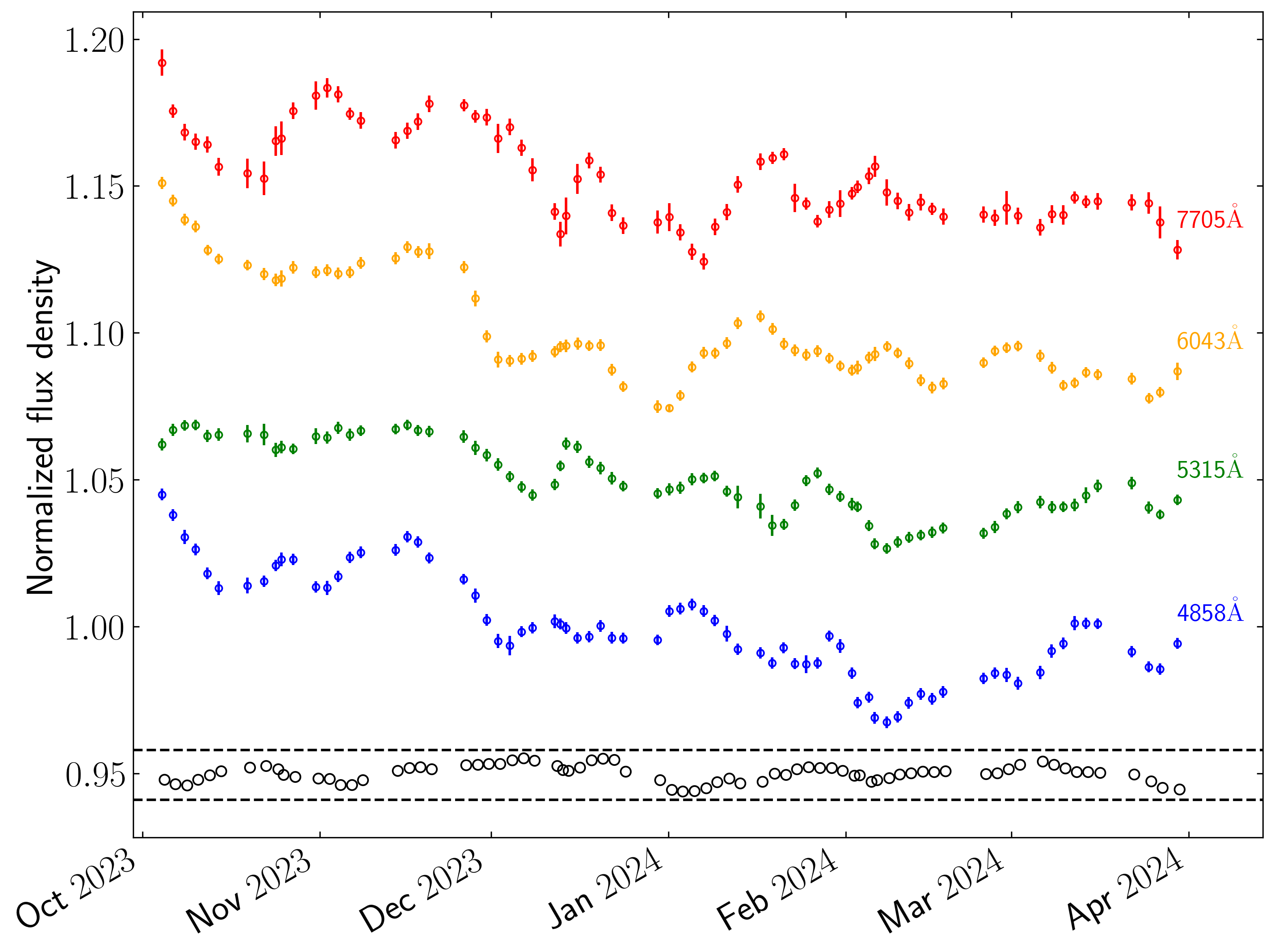}
    \caption{Multi-band light curves of QSO J0455-4216. Variability of around 7\% is observed across all wavelengths for the entire campaign (6 months), while a smaller variability of about 5\% is present within individual months of observations. Micro-variability on the order of 2\%-3\% is evident on timescales of days, with delayed features between bands of a few days. For comparison, the light curve of a representative reference star in the field used in constructing the AGN light curves is shown at the bottom. Dotted lines indicate a 1\% variability and all light curves are vertically shifted for clarity.}  
    \label{fig:Figure1}
\end{figure}

\begin{figure}
    \centering
    \includegraphics[width=\columnwidth]{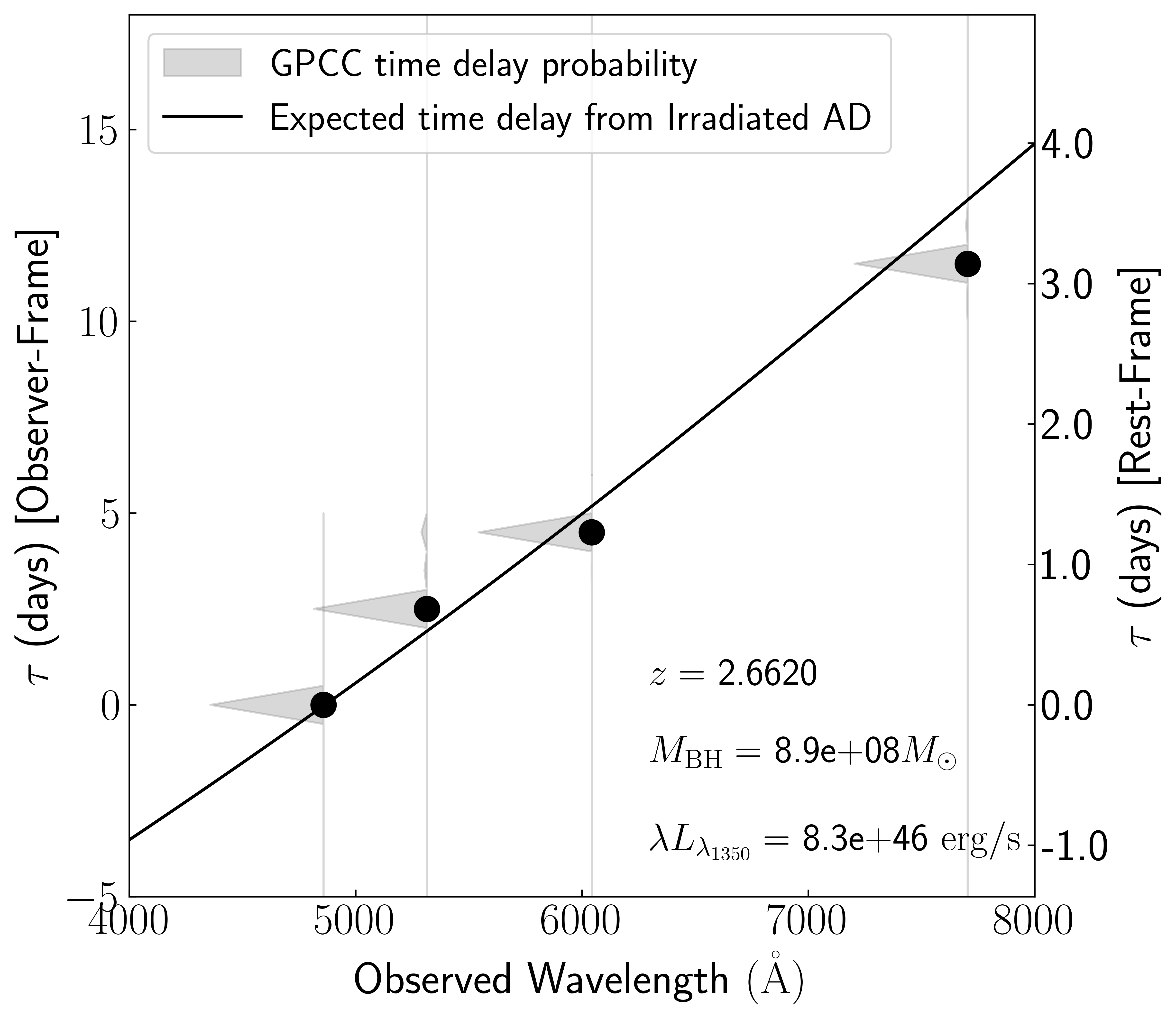}
    \caption{Observed time delay spectrum. The shaded regions represent the time delay ($\tau$) probability distributions obtained using the GPCC method (\citealt{2023A&A...674A..83P}), with black dots indicating the most likely delay corresponding to the peak of each distribution. The solid line denotes the expected delay from an optically thick, geometrically thin accretion disc irradiated by a lamp post geometry. A summary of the physical parameters used in the model, including the black hole mass ($M_{\rm BH}$), luminosity ($\lambda L_{\lambda_{1350}}$), and redshift ($z$), is provided.}  
    \label{fig:Figure2}
\end{figure}

The observed light curves (Fig.~\ref{fig:Figure1}) exhibit a variability with an amplitude of about 7\% and typical uncertainty measurements of 0.2\%.
Given the high luminosity and redshift of this quasar, its short-timescale variability may be classified as micro-variability.
Furthermore, the light curves exhibit highly correlated features, with a Pearson correlation coefficient of 0.9. 
This strong correlation suggests that the variability originates from regions with similar physical conditions, with the primary distinction being the time delay between them. 
The measured time delays to the reference wavelength $4858$\,\AA\, are shown in Fig.~\ref{fig:Figure2}. 
The delay spectrum follows a trend consistent with predictions derived from a SS73 thin accretion disc model irradiated by a so-called lamp-post geometry (\citealt{2007MNRAS.380..669C,2019ApJ...870..123E,2023MNRAS.526..138K,2023MNRAS.522.2002P}). 
Additional delay measurements obtained using various methods (Fig.~\ref{fig:AP2}) yielded a comparable trend, underscoring the robustness of the observed delay spectrum.

\subsection{Modelling the accretion disc response}

The high quality of the light curves and the distinct variability features enabled us to examine the temporal structure of the emission and the origins of the observed delays in unprecedented detail. While variability is often attributed to the disc’s reprocessing layer, recent studies have shown that optical continuum time delays can be significantly affected by diffuse continuum emission (DCE) from BLR clouds (\citealt{2019MNRAS.489.5284K,2019NatAs...3..251C,2022MNRAS.509.2637N}). Thus, even though the delayed emission largely reflects accretion disc reprocessing, a non-negligible DCE component could bias the inferred lag spectrum if left unaccounted for.
To disentangle these effects, we reconstructed the transfer functions associated with both the inner accretion disc and the extended reprocessing region of the BLR. In doing so, we explicitly considered the contributions from DCE and line emission. This approach allows us to test whether the observed variability can be fully explained by the accretion disc or if a fraction of the emission at different wavelengths arises from diffuse, extended BLR gas.

\begin{figure}
    \centering
    \includegraphics[width=\columnwidth]{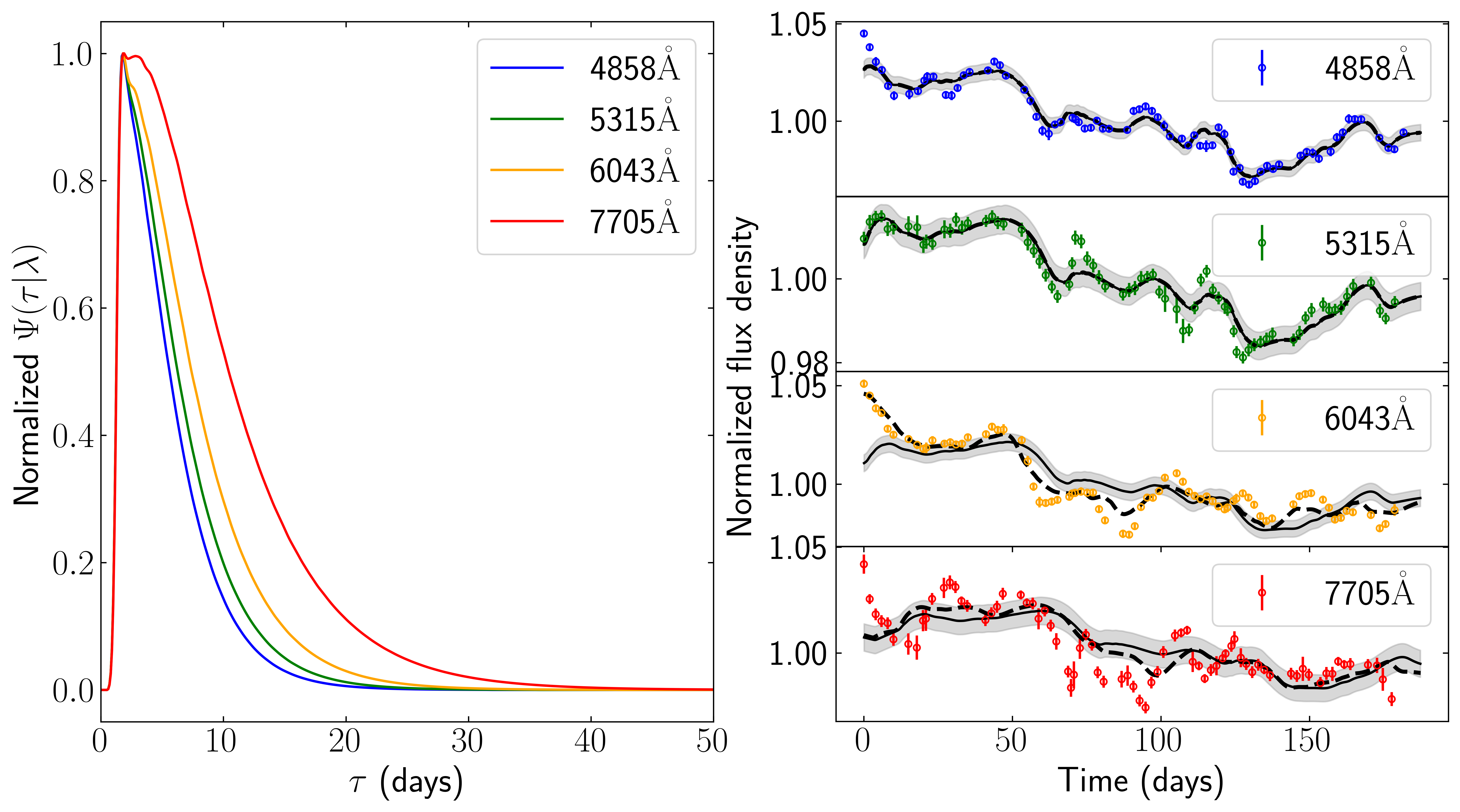}
    \caption{ \textit{Left}: Normalized accretion disc transfer functions \(\Psi(\tau|\lambda)\) for the four photometric bands: \( 4858 \, \text{\AA} \) (blue), \( 5315 \, \text{\AA} \) (green), \( 6043 \, \text{\AA} \) (orange), and \( 7705 \, \text{\AA} \) (red). As predicted by standard thin-disc (SS73) and lamp-post models, the transfer functions broaden with increasing wavelength. Longer-wavelength emission arises from larger radii, resulting in longer time delays due to the greater light travel distance. \textit{Right}: Observed normalized continuum light curves (points with error bars) compared with model predictions. Dashed lines show the results from a single-band modelling approach, in which a latent signal is inferred independently for each filter. Solid black lines show the model light curves computed from the joint GPCC-inferred latent signal, convolved with the corresponding transfer functions. The shaded regions represent the \( 1\sigma \) uncertainties propagated through the model scaling process; see Appendix \ref{ad:simul} for details.}
    \label{fig:Figure3}
\end{figure}

To interpret the wavelength-dependent continuum time delays, we adopted the standard SS73 thin accretion disc model combined with a lamp-post geometry, in which a compact corona illuminates a geometrically thin, optically thick disc. A detailed description of the model implementation is provided in Appendix~\ref{ad:simul}. We used the black hole mass, \( M_{\text{BH}} = (8.9 \pm 1.8) \times 10^{8} \, M_{\odot} \), derived from the CIV line velocity dispersion of \( 3865 \pm 176 \, \mathrm{km\,s^{-1}} \) measured in the SOAR spectrum, employing the single-epoch scaling relation from \cite{2006ApJ...641..689V}. This value is consistent, within the uncertainties, with the SE mass of \( 1.88 \times 10^{9} \, M_{\odot} \) reported by \cite{2018ApJ...865...56L}.
The bolometric luminosity and, consequently, the accretion rate, \( \dot{M} \), was estimated using the rest-frame \( 1350\,\text{\AA} \) continuum, traced by our observed \( 4858\,\text{\AA} \) band. Specifically, we adopted the bolometric correction, \( L_{\mathrm{Bol}} = 3.81 \times \lambda L_{\lambda}(1350\,\text{\AA}) \), following \citet{2024ApJS..272...11P}. The monochromatic luminosity, \( \lambda L_{\lambda} \) at \( 1350 \, \text{\AA} \) is \( 8.25 \times 10^{46} \, \mathrm{erg \, s^{-1}} \), which shows no significant difference from the \( 7.59 \times 10^{46} \, \mathrm{erg \, s^{-1}} \) reported by \cite{2018ApJ...865...56L} nearly a decade ago.

The accretion disc transfer functions are shown in Fig.~\ref{fig:Figure3}, with the corresponding model's light curves compared to the observed continuum variability. Our aim is not to provide a best-fit model for the light curves, as we kept the physical parameters fixed and we did not optimize them. Instead, we focus on demonstrating how well the models  resemble the observed light curves in a qualitative way. The model light curves are computed by convolving disc transfer functions with latent driving signals. We explore two approaches: a single-band approach, where the latent signal is inferred independently for each filter, and a joint analysis, where a single latent signal is reconstructed simultaneously using all filters via the GPCC method. The single-band modelling is particularly useful in practical scenarios where only a few filters may be available. In contrast, the joint approach offers a more physically consistent reconstruction of the common ionizing source. More details are given in Appendix \ref{ad:simul}. 

While the models closely match the observed light curves at shorter wavelengths, a noticeable discrepancy arises in the 7705\AA\ band. The model at this wavelength appears more smeared out compared to the sharper variability features seen in the observations. This implies that the 7705\AA\ emission may not be dominated by the outermost disc regions as anticipated. This discrepancy can be attributed to the broader transfer function at longer wavelengths, which naturally smooths the signal as the light-travel time delays span a wider range of radii in the outer disc. If, however, we artificially narrow the transfer function for the 7705\AA\ band to resemble those of the shorter wavelengths, the model reproduces the observed variability more accurately (not shown). This suggests that the emission contributing to the 7705\AA\ band might not strictly originate from the outermost regions of the accretion disc, as predicted by the standard SS73 theory. Instead, it could trace contributions from regions closer to the inner disc, where the variability remains sharper. This scenario could also explain the deviation seen for the 7705\AA\ band in the time delay spectrum shown in Fig. \ref{fig:Figure2}.

A non-negligible contribution from DCE, which is expected to originate in the BLR, could affect the delay measurements. To assess this possibility, we performed photoionisation simulations to quantify the wavelength-dependent contribution of DCE to the observed flux. In modelling the temporal variability, we account for this inferred DCE by assuming a BLR composed of clouds in Keplerian orbits within a disc-like geometry (\citealt{2023MNRAS.522.2002P}). The DCE contribution to the variability is incorporated as an additional convolved component, with its fractional contribution derived from our photoionisation simulations (see Appendix \ref{dce:simul} for details). 

Figure~\ref{fig:figureSM5} illustrates the effect of DCE on the observed variability and highlights how its inclusion modifies the overall light curve compared to a pure accretion disc model. In the left panel, we show the DCE transfer function, assuming a centroid corresponding to the mean emissivity radius of the CIV BLR as obtained by \cite{2018ApJ...865...56L}, which is 162 light days. For comparison, the inset shows the much narrower AD transfer function for the 5315\text{\AA} band, which reflects the light-travel time delays across the AD. The sharp difference in the widths of the two transfer functions highlights the inherently longer delays and broader smoothing effects associated with the BLR DCE component compared to the accretion disc. In the right panel, we compare the observed light curve to several model scenarios. The blue curve shows the model light curve considering only the AD contribution, demonstrating the expected variability amplitude and structure from the AD transfer function alone. A modest DCE contribution slightly smooths the variability amplitude, but still qualitatively matches the observed light curve. For illustrative purposes, we also included models with higher DCE contributions. As the fractional contribution of DCE increases, the amplitude becomes progressively damped and the light curve is smoothed, clearly departing from the variability patterns predicted by the accretion disc alone. 
A 50\% DCE contribution already produces a significant smoothing effect, which reduces the variability amplitude far beyond what is seen in the 7705\text{\AA} light curve. This indicates that while the AD model captures the overall variability trends at shorter wavelengths, it does not fully account for the sharper features observed at longer wavelengths. Consequently, an additional mechanism or emission component must be invoked to explain the variability structure at 7705\text{\AA}. Nevertheless, the overall agreement between the observed variability patterns and the accretion disc model provides compelling evidence that, for the first time, the continuum time delays at this high redshift are predominantly governed by the accretion disc itself.

\begin{figure}
\includegraphics[width=\columnwidth]{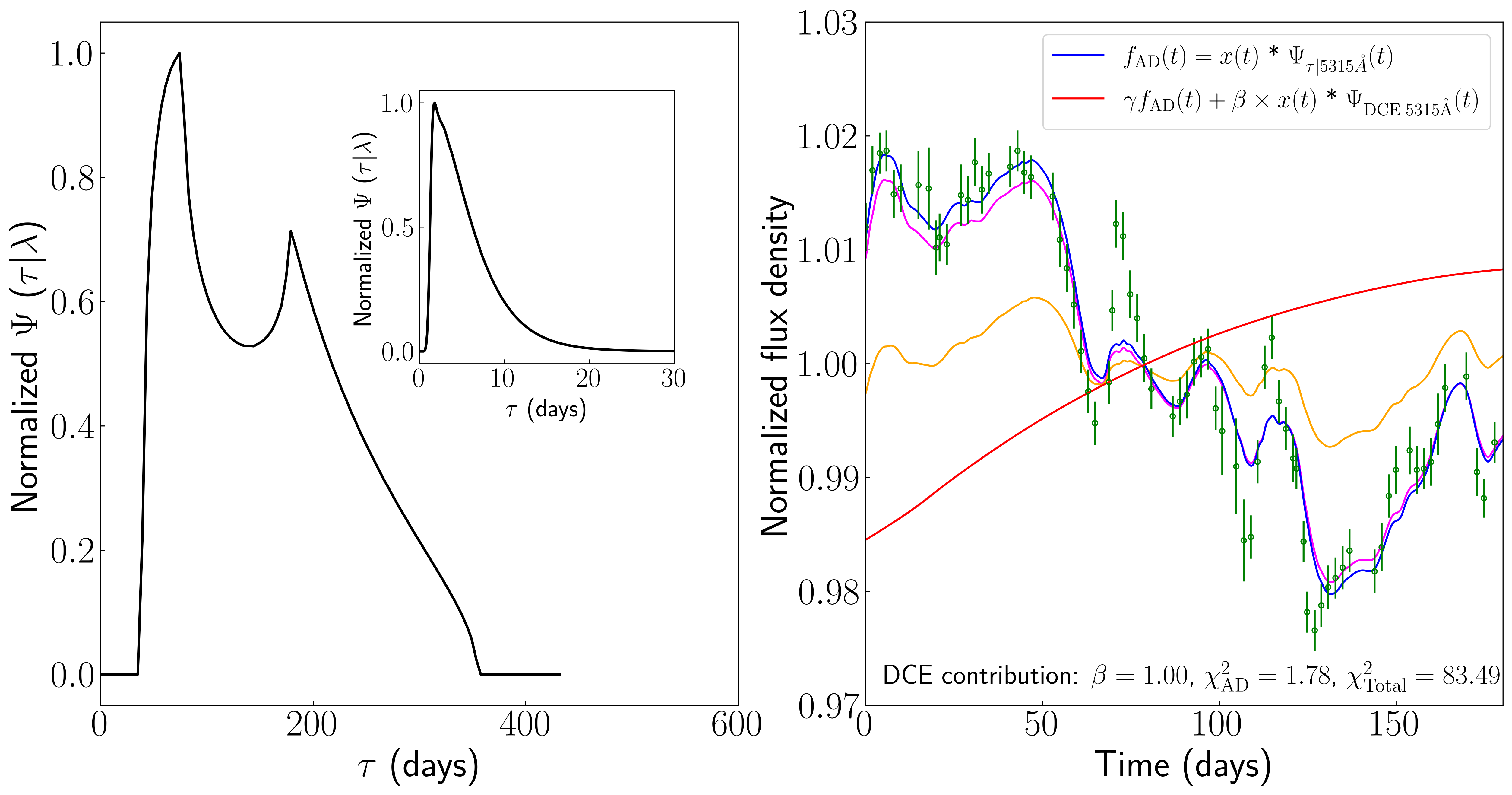}
\caption{\textit{Left}:  DCE transfer function, centred at 162 light days based on the mean emissivity radius of the CIV BLR, is significantly broader than the accretion disc transfer function at \( 5315 \, \text{\AA} \) shown in the inset. Its pronounced double-peaked structure is expected from a disc-like BLR geometry (\( i = 40^\circ \)), resulting in longer and more complex light-travel delays than in the compact inner accretion disc region. \textit{Right}: Observed light curve at \( 5315 \, \text{\AA} \) (green points)  compared to various models. The pure accretion disc model (\( i = 0^\circ \); blue) reproduces variability consistent with the expected broadening from the disc transfer function alone. Introducing a moderate 3.9\% DCE contribution (magenta) slightly smooths the variability amplitude. Increasing the DCE fraction to 50\% (orange) or 100\% (red) overly smooths the light curve, failing to replicate its sharp variability features; see Appendix \ref{sect:model} for details.
}
\label{fig:figureSM5}
\end{figure}

From the measured delay spectrum (Fig.~\ref{fig:Figure2}), we determined a mean responsivity radius of the accretion disc of \( 4.75^{+1.12}_{-1.05} \) light-days in the observer's frame. This provides the first direct measurement of the accretion disc size for a quasar at \( z = 2.66 \). Comparing this result with the CIV BLR mean responsivity radius of \( 592^{+122}_{-38} \) light-days obtained from reverberation mapping (\citealt{2018ApJ...865...56L}), we found that the BLR is approximately 125 times larger than the accretion disc. This ratio is consistent with the theoretically predicted value of 150, assuming the chosen continuum wavelength is closer to the CIV emission line rather than to the disc's mean emissivity radius (\citealt{2024ApJ...968L..16P}). We note that in \citet{2024ApJ...968L..16P}, disc sizes were modelled using a thermal reprocessing framework and recovered through simulated photometric monitoring under realistic observational conditions. By comparing these recovered disc sizes to observed CIV BLR lags, we derived a size–size relation that incorporates luminosity and redshift dependencies, thereby providing a theoretical basis for the observed separation between disc and BLR reverberation lags.

This scaling relationship is particularly significant as it establishes a direct link between the accretion disc size, which can be measured efficiently over a six-month photometric monitoring campaign, and the BLR size. By coupling this result with single-epoch CIV line velocity dispersion measurements, the black hole mass can be estimated without the need for decades-long spectroscopic monitoring of the BLR. This breakthrough highlights the potential for accretion disc size measurements to serve as a faster and more efficient proxy for BLR-based black hole mass determinations, particularly for high-redshift quasars, where monitoring timescales are otherwise prohibitive. However, to determine whether this scaling factor of \(\sim125\ - 150\) is universal across different quasars, it will be essential to monitor a larger sample of sources at high redshifts. Such an effort will allow us to test the robustness of this relationship and explore any dependencies on black hole mass, accretion rate, or quasar properties. Future time-domain surveys and high-cadence monitoring campaigns will be crucial in addressing this question and solidifying the role of accretion disc measurements in black hole mass estimation.

\bigskip
\noindent\textbf{Data availability}\\
\noindent Light curves are available at the CDS via \url{https://cdsarc.cds.unistra.fr/viz-bin/cat/J/A+A/vol/page}

\begin{acknowledgements}

FPN gratefully acknowledges the generous and invaluable support of the Klaus Tschira Foundation. FPN acknowledges funding from the European Research Council (ERC) under the European Union's Horizon 2020 research and innovation program (grant agreement No 951549). SP is supported by the international Gemini Observatory, a program of NSF NOIRLab, which is managed by the Association of Universities for Research in Astronomy (AURA) under a cooperative agreement with the U.S. National Science Foundation, on behalf of the Gemini partnership of Argentina, Brazil, Canada, Chile, the Republic of Korea, and the United States of America. The authors are grateful to Konstantina Boutsia for her assistance with the SOAR reduction.
We thank the anonymous referee for the careful reading of the manuscript and valuable suggestions, which significantly improved the quality and presentation of this work.
This research has made use of the NASA/IPAC Extragalactic Database (NED) which is operated by the Jet Propulsion Laboratory, California Institute of Technology, under contract with the National Aeronautics and Space Administration. This research has made use of the SIMBAD database, operated at CDS, Strasbourg, France.  
      
\end{acknowledgements}

\bibliographystyle{aa}
\bibliography{aa_fp_highz_letter}

\begin{appendix}

\section{Data reduction}
\label{sect:data}

\subsection{WFI images}

The 2.2-meter telescope is equipped with the Wide Field Imager (WFI) camera, which has a field of view of $34'\times33'$, recorded by eight CCDs, each providing a pixel scale of $0.238^{\prime\prime}$/pixel.
The WFI data reduction was carried out using standard IRAF packages alongside custom-developed routines, combined with tools from SExtractor (\citealt{1996A&AS..117..393B}) and Astrometry.net (\citealt{2010AJ....139.1782L}). These were integrated into an automated pipeline designed for accurate source detection and image quality assessment, following the detailed recipes outlined in \cite{2017PASP..129i4101P}.

An analysis of different aperture sizes was conducted to account for the dependence on PSF quality, while minimizing both the host-galaxy contribution and sky background contamination. 
We found that an aperture radius of $1.5^{\prime\prime}$ around the nucleus optimizes the signal-to-noise ratio and results in the lowest absolute scatter in the flux measurements.
For comparison, we also applied image subtraction techniques using the algorithms implemented in the ISIS package (\citealt{2000A&AS..144..363A}). 
This method involves constructing a reference frame from the coaddition of high-quality images, matching the PSF of individual frames with a spatially variable kernel, and subtracting the convolved reference frame from individual images to isolate the AGN variable flux. 
While image subtraction achieves a similar photometric precision of 0.2–0.5 per cent, its performance is sensitive to the quality of the PSF kernel, which depends on the density of stars in the field (\citealt{2017PASP..129i4101P}). 
The field of QSO J0455-4216, with $\sim20$ stars, has fewer stars than more crowded fields (e.g. $\sim1000$ in local Seyfert galaxies), resulting in slightly reduced precision. 
Given the minimal differences in performance and for the sake of clarity, only the light curves derived from aperture photometry are presented in Fig.~\ref{fig:Figure1}.

The absolute flux calibration was performed using non-variable stars flagged for high quality from the DES Data Release 2 catalogue, located in the same field as QSO J0455-4216. 
The fluxes were corrected for atmospheric extinction at the La Silla Observatory (\citealt{1995A&AS..112..383B}) and for Galactic foreground extinction using the recalibrated dust extinction maps from \cite{2011ApJ...737..103S}, adopting $E(B\!-\!V) = 0.044$ as obtained from the NED extinction calculator for the line of sight to QSO J0455-4216.

\subsection{Assessment of photometric systematics}
\label{appendix:photometric_stability}

\begin{figure*}[htbp]
    \centering
    \includegraphics[height=0.5\textheight,width=0.49\textwidth]{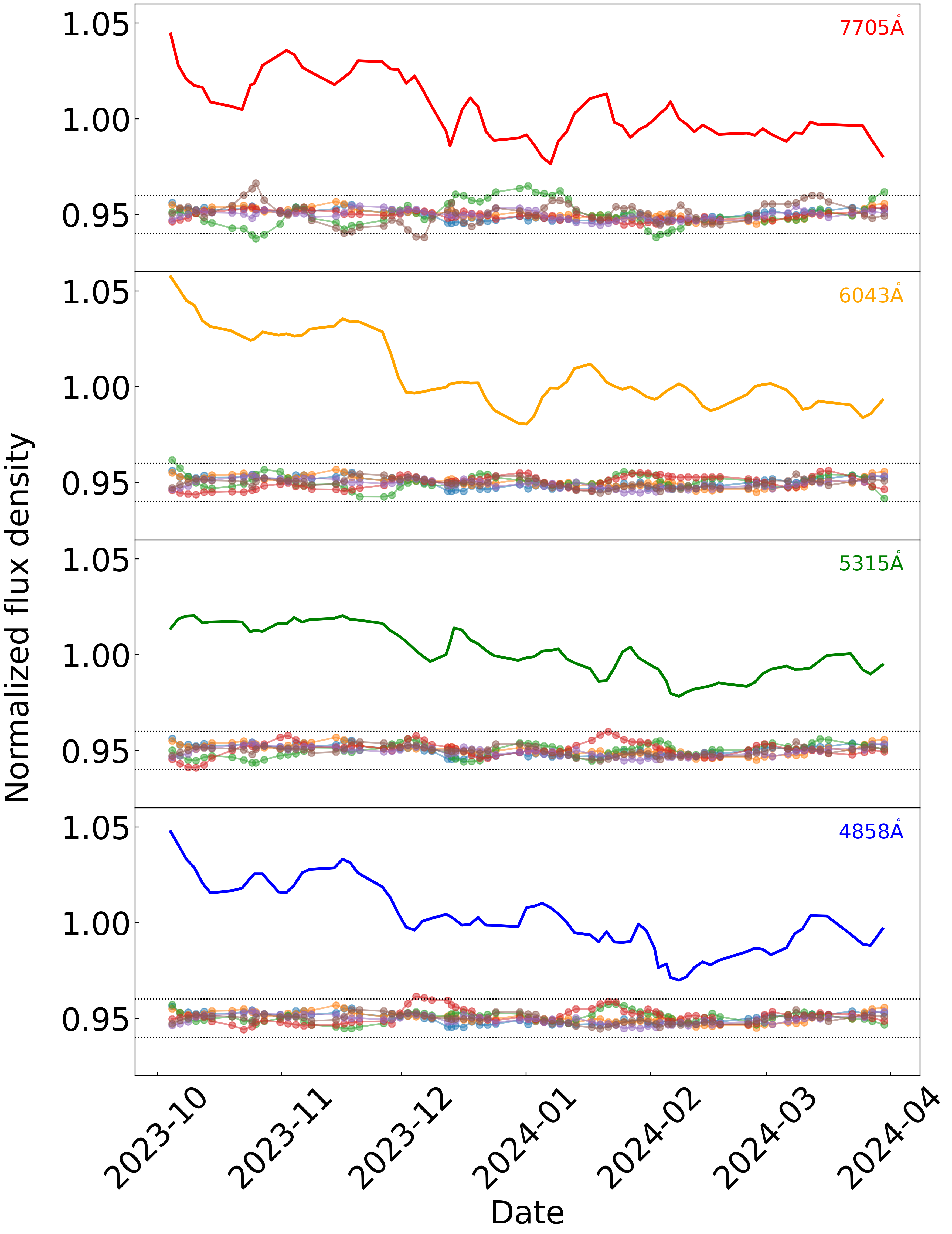}
    \includegraphics[height=0.5\textheight,width=0.49\textwidth]{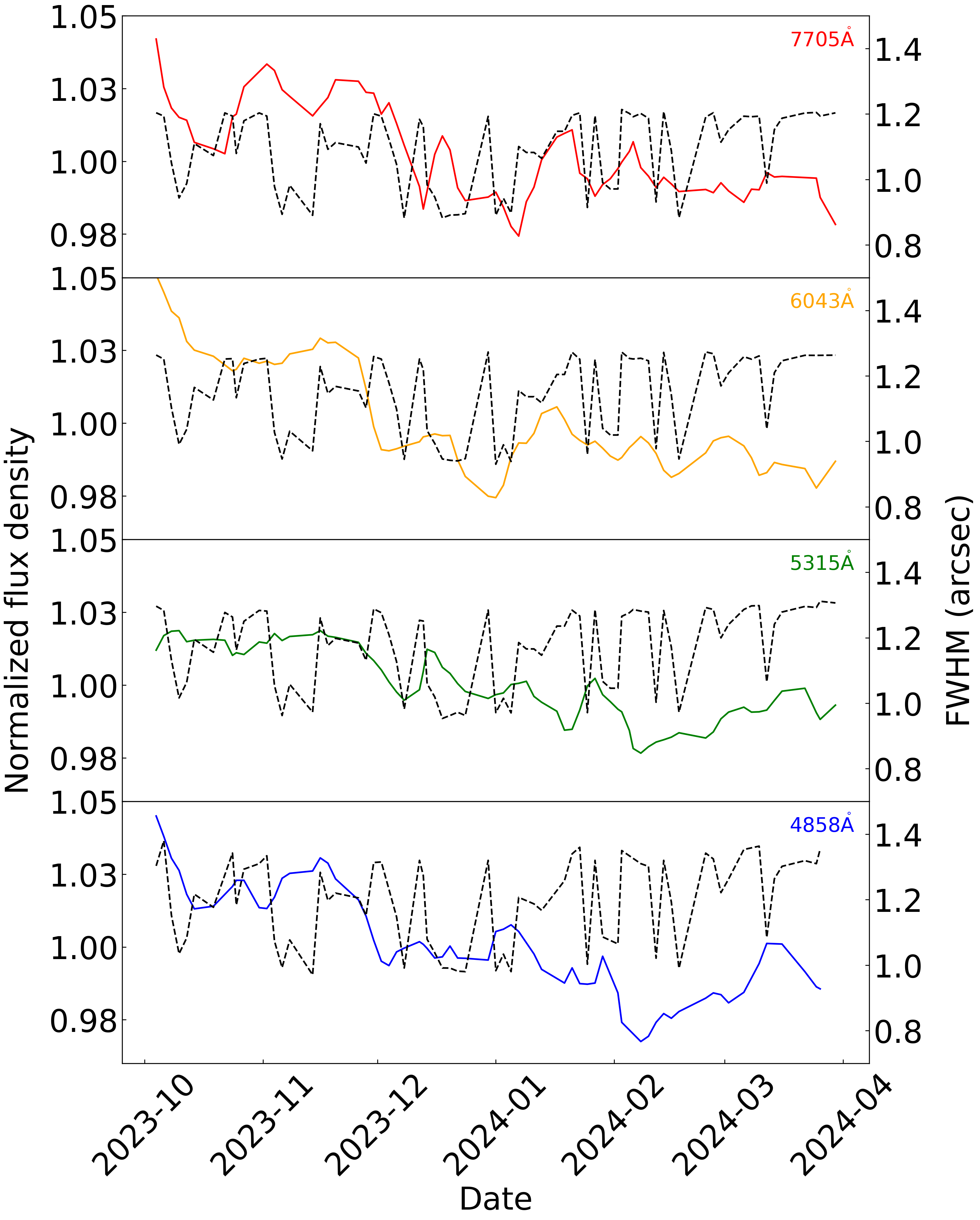}
    \caption{Left: Normalized light curves of the AGN (colored lines) and the six reference stars (colored dots) for all filters, ordered from reddest (top:7705$\,\mathring{\rm{A}}$) to bluest (bottom:4858$\,\mathring{\rm{A}}$). The reference stars exhibit photometric stability throughout the campaign, supporting the robustness of the AGN variability measurement. Right: Comparison of AGN normalized light curves (colored solid lines) with seeing (FWHM, black dashed line) for each filter. No significant correlation is observed between the two quantities in any filter.}
    \label{fig:agn_combined_panels}
\end{figure*}

To ensure the reliability of the AGN light curves and rule out the possibility that the observed variability is driven by systematic effects, we conducted a detailed analysis of potential sources of photometric error. This includes evaluating the behavior of the reference stars, the stability of the seeing, and the potential impact of PSF variations on aperture photometry.
Figure~\ref{fig:agn_combined_panels} (left) shows the normalized light curves of the AGN and the six reference stars in all four filters used for monitoring: $4858\mathring{\rm{A}}$, $5315\mathring{\rm{A}}$, $6043\mathring{\rm{A}}$, and $7705\mathring{\rm{A}}$. The AGN exhibits clear and significant variability on timescales of weeks, with amplitudes between 5\% and 8\%. In contrast, the reference stars remain photometrically stable, with root-mean-square (RMS) fluctuations typically in the range 0.2--0.5\%, and maximum deviations below 1\%.

Across all filters, the AGN variability is substantially larger than that of the reference stars. The latter show no long-term trends or coherent low-frequency variations, consistent with the absence of red noise. Pairwise Pearson correlation coefficients computed among all reference stars yielded negligible values (e.g. $r = -0.004$, $p = 0.97$), indicating that the stars do not share common variability patterns that could hint at systematic errors.

To explore whether seeing variations could affect the photometry, we examined the correlation between AGN flux and seeing (FWHM) for each band. Figure~\ref{fig:agn_combined_panels} (right) shows the AGN light curves overlaid with the corresponding seeing measurements (dashed lines). The computed Pearson correlation coefficients are
\begin{itemize}
\item Filter 860: $r = -0.130$, $p = 0.296$,
\item Filter 875: $r = -0.213$, $p = 0.067$,
\item Filter 864: $r = 0.043$, $p = 0.719$,
\item Filter 849: $r = 0.057$, $p = 0.627$.
\end{itemize}

These results indicate weak and statistically insignificant correlations between seeing and AGN flux in all filters ($|r| < 0.22$, $p > 0.05$). Moreover, the median seeing values in all bands are comfortably below the adopted aperture diameter of $3^{\prime\prime}$ ($1.5^{\prime\prime}$ radius): $1.217^{\prime\prime}$ ($4858\mathring{\rm{A}}$), $1.20^{\prime\prime}$ ($5315\mathring{\rm{A}}$), $1.17^{\prime\prime}$ ($6043\mathring{\rm{A}}$), and $1.11^{\prime\prime}$ ($7705\mathring{\rm{A}}$). This confirms that the photometric aperture captures the full AGN flux without significant seeing-related losses.

\begin{figure}
    \centering
    \includegraphics[width=\columnwidth]{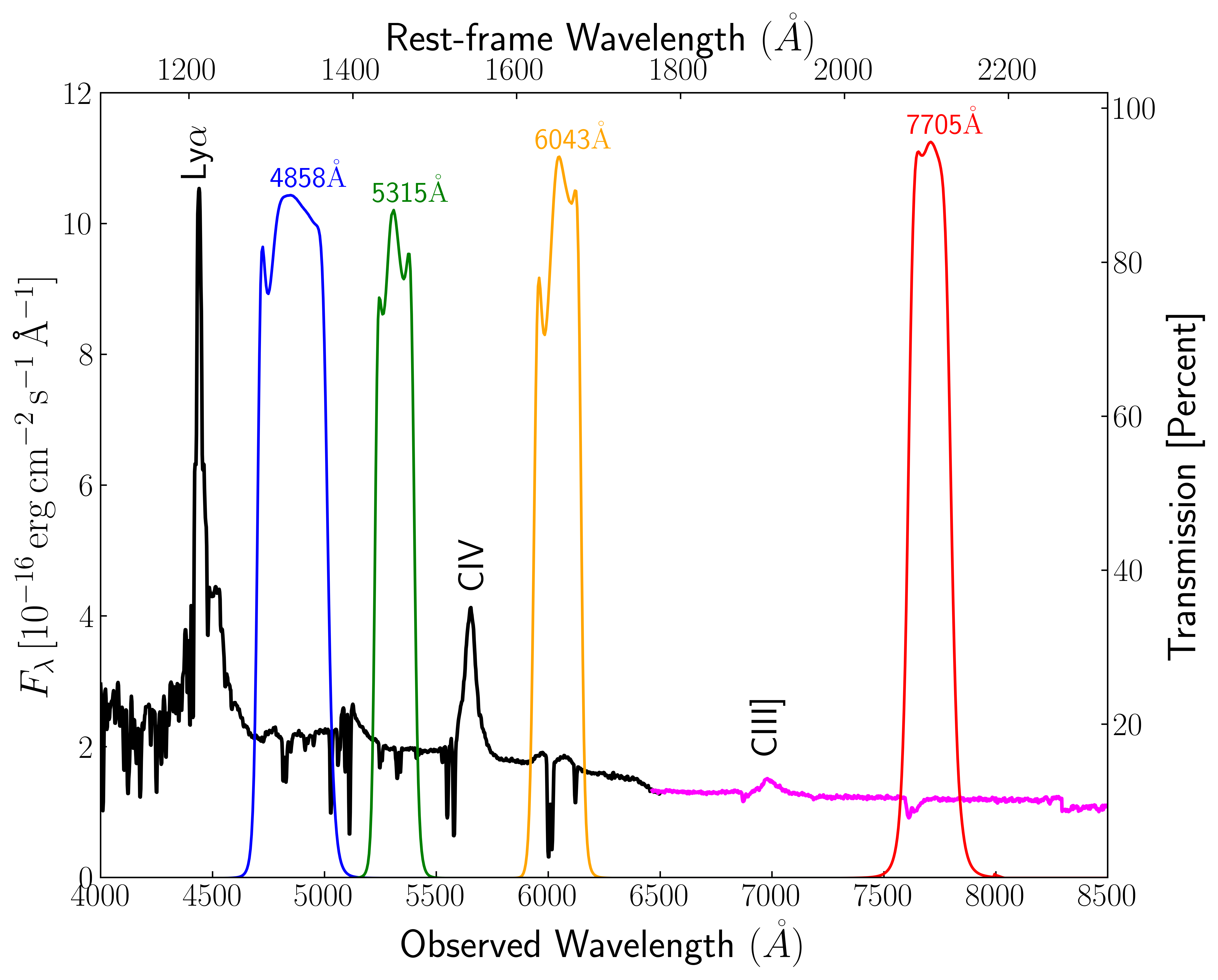}
    \caption{Effective transmission of the medium-band filters used in the monitoring. The coloured lines show the transmission profiles of the WFI medium-band filters. These profiles represent the measured filter transmission functions and do not include convolution with the quantum efficiency of the WFI camera.  Overlaid in black is the spectrum of QSO J0455-4216 observed with SOAR, with relevant emission lines indicated. These filters predominantly capture the AGN continuum variations free of strong emission lines near the \civ{} and \ciii{} features. In particular, the $4858$\AA\ band measures the rest-frame flux at $1350$\AA\ and is used as a reference to determine time delays. The $6043$\AA\ filter includes the HeII$\lambda1640$ emission line and a local absorption feature; however, together they contribute less than 2\% of the integrated flux in this band and are not expected to significantly affect the measured continuum time delays.
    The portion of the spectrum extending beyond the maximum coverage of Goodman Spectrograph in its Blue configuration ($> 6500\text{\AA}$) is shown in magenta and taken from \cite{2018ApJ...865...56L} (private communication).} 
    \label{fig:AP1}
\end{figure}

\subsection{SOAR spectrum}

The optical spectrum of QSO J0455-4216 was observed using the 4.1m Southern Astrophysical Research (SOAR\footnote{Based in part on observations obtained at the Southern Astrophysical Research (SOAR) telescope, which is a joint project of the Minist\'{e}rio da Ci\^{e}ncia, Tecnologia e Inova\c{c}\~{o}es (MCTI/LNA) do Brasil, the US National Science Foundation’s NOIRLab, the University of North Carolina at Chapel Hill (UNC), and Michigan State University (MSU).}) Telescope on February 14, 2024 (Proposal Code: SO2024A-013 -- PI: Panda).
The observations were performed with the Goodman Spectrograph (\citealt{2004SPIE.5492..331C}) in its Blue Camera (M1 mode) configuration.
We employed the 400 SYZY grating, providing a resolving power of about 1000 at 5000$\text{\AA}$, covering a wavelength range of $3031 - 6500\text{\AA}$. 
The data were obtained in spectroscopic mode with a long slit of width $1.0^{\prime\prime}$, in three exposures of 15 minutes each.
The detector was operated with a read noise of $3.69\ e^-$ and a gain of $0.56\ e^-/\mathrm{ADU}$ with a $1 \times 1$ binning mode, yielding high spatial resolution.
The observation was conducted at an average airmass of 1.32, with atmospheric conditions of approximately $19^\circ$C and $23.5\%$ relative humidity.
The data reduction was carried out using the Goodman Spectroscopic Pipeline\footnote{\url{https://soardocs.readthedocs.io/projects/goodman-pipeline/en/latest/overview.html}} (version 1.3.8).
After the initial bias subtraction, flat-field correction, and cosmic ray rejection, we performed the wavelength calibration using a polynomial fit to the arc lamp spectrum, achieving an RMS accuracy of $0.512$\AA. Sky subtraction, 1D extraction, and flux calibration were subsequently applied. The final spectrum, extracted from an aperture spanning $1.7738^{\prime\prime}$, was flux-calibrated using standard star observations and corrected for telluric absorption using combined frames.
Wavelength calibration was performed using a comparison lamp (HgArNe) exposure taken immediately after the target observation. Figure~\ref{fig:AP1} shows the effective transmission and position of the medium-band filters together with the SOAR spectrum of QSO J0455-4216.

\section{Time series analysis}

To estimate the time delays between the continuum bands, we applied five different methods: the interpolated cross-correlation function (ICCF), the Z-transformed discrete correlation function (ZDCF), the Von-Neumann (VN) estimator, JAVELIN, and the Gaussian Process Cross-Correlation (GPCC) method. In all cases, we used the shortest-wavelength band as the reference light curve.

The \textit{ICCF} (\citealt{1987ApJS...65....1G}) method measures time delays by cross-correlating two unevenly sampled light curves. One light curve is linearly interpolated while the other remains unaltered and this process is then reversed. The cross-correlation function, \( R(\tau) \), is computed over a range of lags, \(\tau\). We determined the time delay as the centroid of the cross-correlation function above a threshold of \( R \geq 0.8R_{\rm max} \), where \( R_{\rm max} \) is the peak correlation coefficient. To avoid artifacts caused by the linear interpolation of the data, we used an interpolation step of 2 days which corresponds to the mean sampling cadence of the light curves.

The \textit{ZDCF} (\citealt{1997ASSL..218..163A}) method improves upon the traditional discrete correlation function (DCF) by addressing uneven sampling more effectively. Instead of binning time-delay pairs uniformly, ZDCF uses equal population bins, which improves the statistical robustness of the correlation estimates. Fisher’s \( z \)-transform is applied to derive confidence levels for the correlation coefficients. As with the ICCF, the time lag was determined using the centroid of the ZDCF above \( 0.8R_{\rm max} \).

To complement the cross-correlation methods, we also employed the Von-Neumann (VN) estimator (\citealt{2017ApJ...844..146C}), a statistical estimator that avoids interpolation and binning by quantifying the degree of randomness within a combined light curve. It computes a statistic $\mathcal{V}_N(\tau)$ defined as 
\begin{equation}
\mathcal{V_{N}(\tau)} = \frac{1}{N-1} \sum_{i=1}^{N-1} \frac{\left[ F(t_i) - F(t_{i+1}) \right]^2}{W_{i,i+1}},
\end{equation}
where \( W_{i,i+1} = 1/[\sigma_{lc}^2(t_i) + \sigma_{lc}^2(t_{i+1})] \) accounts for the flux uncertainties \(\sigma_{lc}\). The optimal time delay \(\tau_0\) is obtained as the lag that minimizes \(\mathcal{V_{N}(\tau)}\) over the predefined search range.

For ICCF, ZDCF, and VN, we estimated uncertainties using flux randomization and random subset selection (FR/RSS; \citealt{2004ApJ...613..682P,1999PASP..111.1347W}). We generated 2000 mock light curves by resampling the original data \( N \) times with replacement (bootstrap), such that each mock curve contains \( N \) points with some epochs duplicated and others omitted—resulting in \(\sim\)63\% unique epochs on average. Flux values were perturbed with Gaussian noise based on their measurement uncertainties. We then computed the ICCF and ZDCF centroids, and the VN dispersion statistic, for each mock pair, and estimated uncertainties from the central 68\% range of the resulting lag distributions.

\textit{JAVELIN} (\citealt{2011ApJ...735...80Z}) employs a Bayesian framework, representing the driving light curve as a damped random walk (DRW) and the target light curve as a shifted, scaled, and smoothed response using a top-hat transfer function. Using Markov chain Monte Carlo (MCMC) sampling, JAVELIN obtains posterior distributions for the lag, scaling factor, and DRW parameters. The final lag is the centroid of the posterior distribution, and its width defines the uncertainty. This method naturally incorporates observational noise and irregular sampling.

Finally, we applied the Gaussian process cross-correlation (GPCC) method (\citealt{2023A&A...674A..83P}), a newly developed approach that models the light curves probabilistically using Gaussian processes (GPs). The observed flux in each light curve is expressed as  
\begin{equation}
y_l(t) = \alpha_l f(t - \tau_l) + b_l + \epsilon_l,
\end{equation}
where \( \alpha_l \) and \( b_l \) are scaling and offset parameters, \(\tau_l\) is the time delay, and \(\epsilon_l\) is Gaussian noise. The latent signal \( f(t) \) is modelled as a zero-mean GP with a covariance kernel \( k_{\rho}(t, t') \), which describes the temporal correlations in the variability. GPCC jointly estimates the time delays by maximizing the likelihood of the observed data and delivers posterior probability distributions for the lags. By incorporating observational errors, irregular sampling, and physically motivated priors (e.g. the AD/BLR size-luminosity relation), GPCC ensures a robust and probabilistic estimation of the delays.

\begin{figure}
    \centering
    \includegraphics[width=\columnwidth]{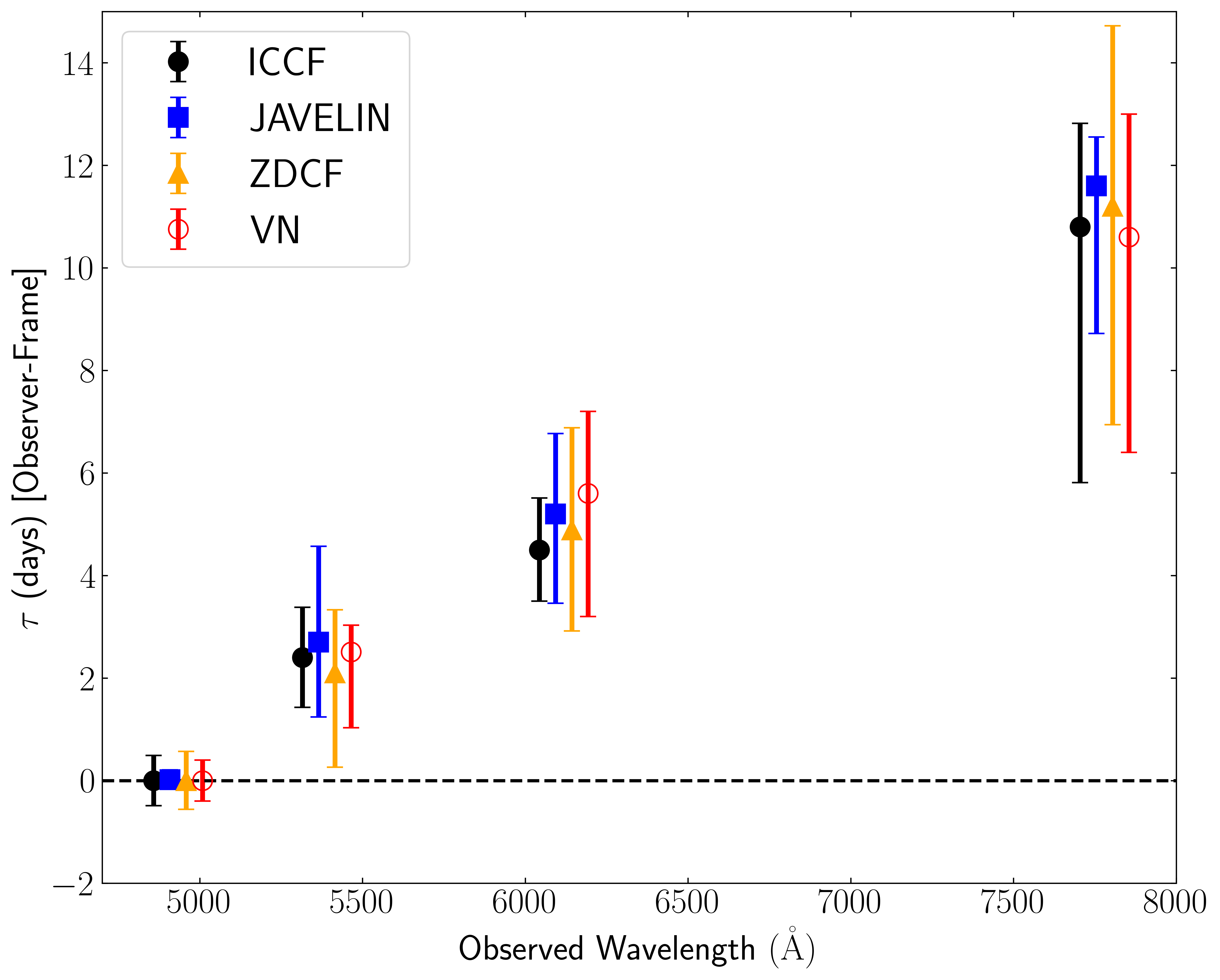}
    \caption{Observed time delay spectrum for multiple delay estimation methods. Time delays were measured using the interpolated cross-correlation function (ICCF;  \citealt{1987ApJS...65....1G} black circles), JAVELIN (\citealt{2011ApJ...735...80Z}) damped random walk modelling (blue squares), the $z$-transformed discrete correlation function (ZDCF; \citealt{1997ASSL..218..163A} yellow triangles), and the von Neumann statistical estimator (VN; \citealt{2017ApJ...844..146C} red open circles), with error bars corresponding to 1$\sigma$ uncertainties. A small shift in wavelength has been introduced for clarity. The close agreement between delays obtained by different methods demonstrates the robustness of the measurements, supporting the reliability of the wavelength-dependent delay structure observed in the quasar's variability.}
    \label{fig:AP2}
\end{figure}

The time delay measurements obtained by these various methods are shown in Fig.~\ref{fig:AP2}. The combination of multiple, independent techniques enables cross-verification, helping us to identify systematic biases and ensuring the robustness of our conclusions.

\begin{table*}
\centering
\renewcommand{\arraystretch}{1.4}
\caption{Time-delay measurements in observer-frame days using ICCF, ZDCF, JAVELIN, VN, and the GPCC model. }
\begin{tabular}{ccccccccc}
\hline\hline
Filter & $\lambda_{\mathrm{obs}}$ [\AA] & Mean Flux [mJy] & $F_{\mathrm{var}}$ & ICCF & JAVELIN & ZDCF & VN & GPCC \\
\hline
860 & 4858 & 0.22 & 0.018 &
$0.00_{-0.49}^{+0.49}$ &
$0.02_{-0.17}^{+0.19}$ &
$0.00_{-0.56}^{+0.57}$ &
$0.00_{-0.40}^{+0.40}$ &
$0.00 \pm\,0.30$ \\
875 & 5315 & 0.13 & 0.012 &
$2.40_{-0.97}^{+0.98}$ &
$2.70_{-1.46}^{+1.87}$ &
$2.10_{-1.84}^{+1.23}$ &
$2.51_{-1.48}^{+0.52}$ &
$2.59 \pm\,0.87$ \\
864 & 6043 & 0.19 & 0.018 &
$4.50_{-1.10}^{+1.11}$ &
$5.20_{-1.74}^{+1.57}$ &
$4.89_{-1.97}^{+1.99}$ &
$5.60_{-2.40}^{+1.60}$ &
$4.75 \pm\,1.08$ \\
849 & 7705 & 0.12 & 0.015 &
$10.80_{-4.99}^{+2.02}$ &
$11.60_{-2.88}^{+0.95}$ &
$11.20_{-4.26}^{+3.52}$ &
$10.60_{-4.20}^{+2.40}$ &
$11.51 \pm\,2.67$ \\
\hline
\end{tabular}
\tablefoot{GPCC delays are reported as the mode of the posterior distribution, with uncertainties corresponding to the 68\% highest posterior density (HPD) interval. The table also includes the mean flux density (in mJy) of the light curves and the fractional variability $F_{\mathrm{var}}$ following \cite{1997ApJS..110....9R}.}
\label{tab:delays_all_methods}
\end{table*}

\section{Modelling}
\label{sect:model}

\subsection{Accretion disc reprocessed light curve}
\label{ad:simul}

To model the wavelength-dependent continuum time delays, we adopted the standard SS73 thin accretion disc framework in combination with a lamp-post geometry, following the formalism described in \citet{2023MNRAS.522.2002P}. In brief, this scenario describes a geometrically thin, optically thick accretion disc extending from the innermost stable circular orbit at $r_{\text{in}} \sim 6R_{g}$ to an outer radius $r_{\text{out}}$, illuminated by a compact X-ray emitting corona located at a fixed height $h$ above the black hole along its rotational axis. In this framework, the local disc temperature reflects the combined contribution of internal viscous heating and external irradiation. While the canonical scaling $T(r)\propto (M_{\rm BH} \dot M)^{1/4}r^{-3/4}$ describes the viscous heating alone, irradiation from the corona elevates the temperature above this level, modifying the overall profile.

The thermal energy radiated at each radius results from a combination of internal viscous heating and external irradiation by the corona. The total emission can be modelled as blackbody radiation integrated over the radial extent of the disc. The observed continuum flux at each wavelength is obtained by convolving the driving X-ray light curve with the accretion disc transfer function.

The variability in the driving continuum is modelled as a stochastic process with a power spectral density $P(\nu) \propto \nu^{-2}$, corresponding to a random walk process (\citealt{2023MNRAS.522.2002P}). The observed continuum flux $F_{c}(\lambda,t)$ at wavelength $\lambda$ is obtained by convolving the driving X-ray light curve $F_{x}(t)$ with the accretion disc transfer function, $\Psi(\tau|\lambda)$, expressed as  
\begin{equation}
F_{c}(\lambda,t) = \int_{0}^{\infty} \Psi(\tau|\lambda) F_{x}(t-\tau) \, \mathrm{d}\tau,
\end{equation}
where the transfer function, $\Psi(\tau|\lambda)$, incorporates the light-travel delays across the disc. The time delay for a radius, $r$, inclination, $i$, and azimuthal angle, $\phi$, in the lamp-post geometry is given by (\citealt{2005ApJ...622..129S}):
\begin{equation}
\tau(r, \phi, i) = \frac{1}{c} \left[ \sqrt{r^2 + h^2} + r \sin i \cos \phi + h \cos i \right],
\end{equation}
where $c$ is the speed of light and $h$ is the height of the corona above the disc. In the following, we assume \( i = 0^\circ \), since the inclination has only a marginal effect on the centroid of the transfer function (\citealt{2023MNRAS.522.2002P}).

To explore how well the disc reprocessing scenario can account for the observed variability at different wavelengths, we implemented two modelling approaches. In the first, each observed AGN light curve was treated independently. For each photometric band, we inferred a latent driving signal directly from the corresponding light curve, without assuming a common origin across bands. This approach remains agnostic about the physical nature of the driver, which is particularly valuable in practical situations where only a small number of bands are available. This methodology is similar to that implemented in RM tools such as JAVELIN, which can also model each light curve independently using a damped random walk as the latent driving process. Moreover, this single-band modelling strategy ensures that the light curves can be analyzed even in reduced-filter settings, without requiring a complete or uniform multi-band dataset. It also avoids introducing biases that could arise from assuming a shared underlying signal, which may be difficult to constrain in the presence of sparse or uneven sampling. The inferred latent signal was then convolved with the corresponding disc response function $\Psi(\tau|\lambda)$ to produce model light curves for each band. These reconstructions are shown as dotted lines in Fig.~\ref{fig:Figure3} (right panel). 

In the second approach, we adopted a joint modelling framework in which a single latent driving signal was inferred from the full set of AGN light curves across all bands simultaneously using the GPCC method. This joint analysis is designed to reconstruct a common underlying variability pattern while accounting for the wavelength-dependent time delays encoded in the disc transfer functions. Figure~\ref{fig:Figure3} (right) shows both the original per-band models (dotted lines) and the updated GPCC-based joint models (solid black lines with shaded uncertainties), allowing a direct comparison between the two approaches.
The shaded regions (``error snakes") in the model light curves represent the $1\sigma$ uncertainties resulting from propagating the observational errors through the model scaling process. Specifically, the model flux is scaled to match the observed light curves using a linear transformation, where the scaling parameters and their uncertainties are determined via curve fitting. The propagated uncertainties on the scaled model flux are computed using the covariance matrix of the fit parameters. These regions provide a visual representation of the uncertainty in the model prediction.
The joint modelling yields particularly good agreement with the light curves at the two shortest wavelengths (4858\AA\ and 5315\AA\ ), comparable to the quality of the fits obtained with the inferred latent signal in the original single-band approach. In contrast, the longer-wavelength bands (6043\AA\ and 7705\AA\ ) show less accurate fits, which may suggest that the outer regions of the accretion disc, where these wavelengths originate, respond differently or are affected by additional processes not fully captured by the standard reverberation model. While the GPCC model ensures physical consistency by enforcing a shared driver, its performance varies across wavelengths, potentially indicating that the emission at longer wavelengths is not fully governed by the same reprocessing mechanism as the shorter ones. This could reflect additional contributions from reprocessing on larger scales, or even wavelength-dependent deviations from the assumed disc geometry. These possibilities are beyond the scope of the current model but merit further investigation in future work.

\subsection{DCE and photoionization simulations}
\label{dce:simul}

\begin{figure}
\includegraphics[width=\columnwidth]{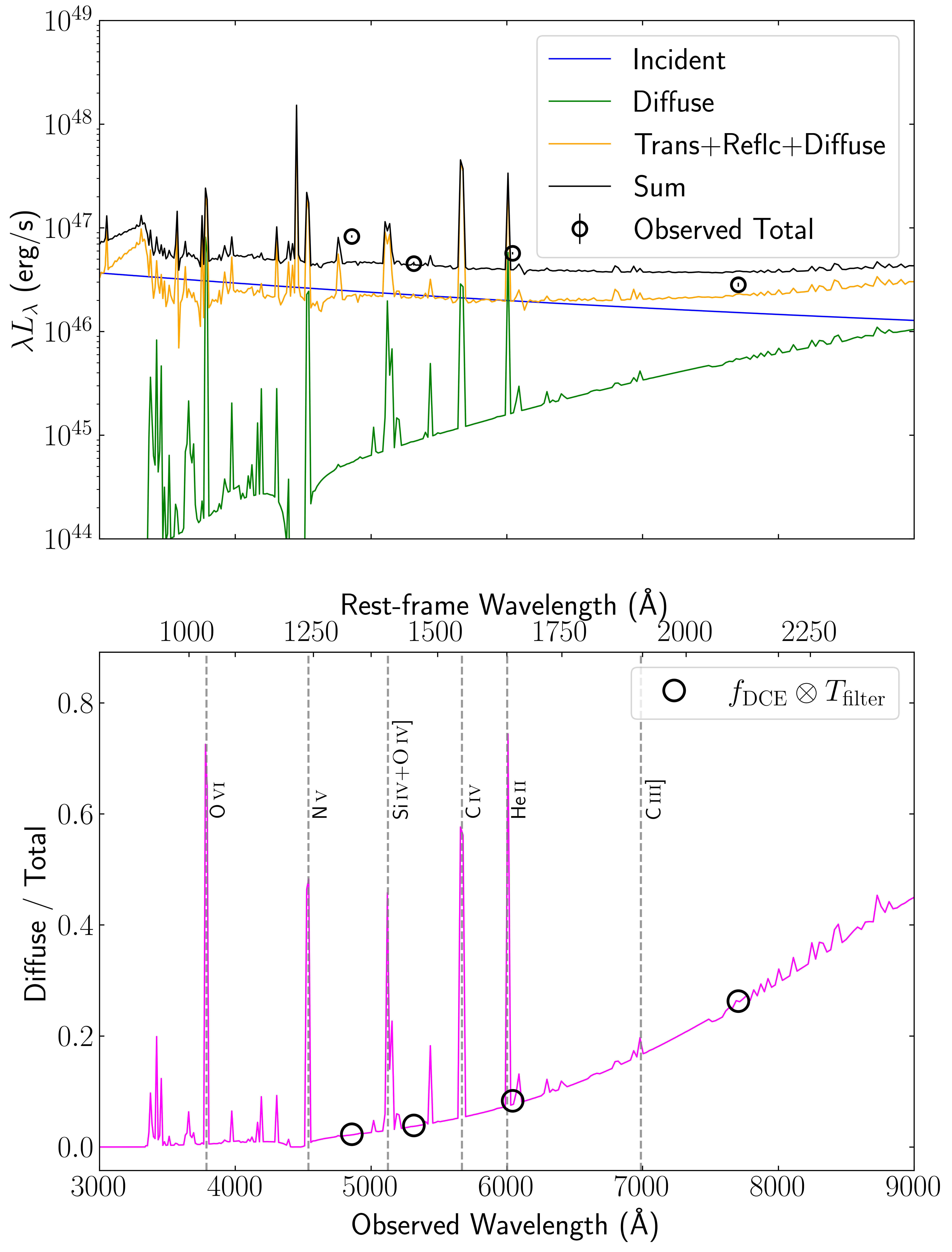}
\caption{Photoionisation simulations and DCE fraction. \textit{Top}:  Luminosity contributions from various components are shown as a function of wavelength. The blue curve represents the incident emission, i.e. the intrinsic continuum irradiating the BLR clouds. The green curve shows the diffuse emission from photoionised gas within the BLR. The orange curve is the sum of transmitted, reflected, and diffuse components.  The black curve represents the total emission model (sum of all components), which is in good qualitative agreement with the observed total emission (black circles) derived from the photometric light curves. \textit{Bottom}:  Fractional contribution of DCE to the total emission is plotted in magenta. Prominent DCE features linked to major BLR emission regions are marked in the bottom panel.
The black circles indicate the effective DCE fractions integrated over the photometric filter bands employed in this study (see text for details).
An accompanying animation is available online, illustrating the full range of DCE contributions and their progressive impact on the light curve.
}
\label{fig:AP3}
\end{figure}

The DCE is primarily composed of hydrogen/helium free-bound continua, free-free emission, and scattering processes including electron and Ly$\alpha$ Rayleigh scattering (\citealt{2001ApJ...553..695K}). Its presence has been observed as an excess in the lag spectrum, most prominently at shorter wavelengths near the Balmer limit and at longer wavelengths near the onset of the Paschen lines, but with contributions predicted across a broad spectral range (1000–10000\AA).
The simulations were carried out using the CLOUDY photoionisation code (version C23.01; \citealt{2023RNAAS...7..246G}), which models the physical conditions of ionized gas and its resulting emission spectra. CLOUDY self-consistently calculates the radiative transfer and emission processes in photoionised clouds under a wide range of conditions. 
In our setup, we assumed a hydrogen density of \( n_{\text{H}} = 10^{12} \, \mathrm{cm^{-3}} \), typical of BLR clouds \citep{2008ApJ...675...83B, 2021ApJ...907...12S}, and located the cloud at a mean emissivity radius of \( 162 \, \text{light-days} \) in the rest-frame, as determined from the reverberation mapping of the CIV line (\citealt{2018ApJ...865...56L}). The model adopts a single-cloud approximation with a column density of \( 10^{23} \, \mathrm{cm^{-2}} \), and explores a range of hydrogen densities (\( 10^{10} - 10^{14} \, \mathrm{cm^{-3}} \)) to determine the best match with the observed spectrum. The covering fraction is set to 20\% \citep{2019MNRAS.489.5284K}.
The incident ionizing continuum was taken as a standard SS73 AGN spectral energy distribution, coupled with a warm comptonising coronal component \citep{2018MNRAS.480.1247K} modelled using the black hole mass and Eddington ratio obtained from the spectral fitting. The incident SS73 AGN continuum illuminates the cloud and produces a combination of transmitted, reflected, and diffuse emissions. This DCE, arising primarily from hydrogen and helium free-bound processes and scattering, was extracted to analyze its fractional contribution to the observed continuum variability. 
Figure~\ref{fig:AP3} (top panel) shows the luminosity contributions from the different components, including the incident emission, which represents the intrinsic continuum irradiating the BLR clouds, and the diffuse emission, arising from photoionised gas within the BLR. These results correspond to a single-cloud photoionisation model with $n_{\text{H}} = 10^{12} \, \mathrm{cm^{-3}}$, column density of $10^{23} \, \mathrm{cm^{-2}}$, solar metallicity, and a fixed distance of 162 light-days. A constant covering fraction of 20\% is assumed, and no additional kinematic broadening is applied beyond CLOUDY's default spectral binning of 1\,\AA. The sum of all components, including transmitted, reflected, and diffuse contributions, is shown alongside the observed total emission, derived from our photometric light curves. There is a good qualitative agreement between the observed and modelled emissions. The bottom panel in Fig.~\ref{fig:AP3} shows the fractional contribution of the DCE, as a function of wavelength. The black circles represent the effective DCE fractions integrated over the photometric filter bands used in this study. The DCE contribution is approximately 2.3\% at \( 4858 \, \text{\AA} \), 3.9\% at \( 5315 \, \text{\AA} \), 8.3\% at \( 6043 \, \text{\AA} \), and 26.3\% at \( 7705 \, \text{\AA} \). These values indicate that the DCE contribution is small at shorter wavelengths but increases significantly toward longer wavelengths.

To account for the inferred DCE in the temporal variability model, we assume the BLR is composed of clouds in Keplerian orbits within a disc-like geometry, following \cite{2023MNRAS.522.2002P}. The BLR is inclined at $40^{\circ}$, and its radial extent is constrained to $R_{\text{max}} / R_{\text{min}} = 2.0$. The DCE contribution to the variability is incorporated as an additional convolved component, with its fractional contribution derived from our photoionisation simulations.

The final total continuum light curve is given as  
\begin{equation}
F_{\text{Total}}(t) = \gamma F_{\text{AD}}(t) + \beta \left[ F_{\text{AD}}(t) * \Psi_{\text{DCE}}(\tau) \right],
\label{ref:eq5}
\end{equation}
where $F_{\text{AD}}(t)$ is the result of convolving the random walk driving light curve $F_{x}(t)$ with the transfer function for the accretion disc, $\Psi_{\text{AD}}(\tau)$, and $\Psi_{\text{DCE}}(\tau)$ is the DCE transfer function. Here, $\gamma = 1 - \beta$, with $\beta$ representing the fractional contribution of DCE derived from our photoionisation simulations. A value of $\beta = 1$ implies a pure DCE-dominated scenario, while $\beta = 0$ corresponds to a pure accretion disc contribution.

The derived DCE fractions (Fig.~\ref{fig:AP3} bottom panel) are incorporated into Eq.~\ref{ref:eq5} as the parameter \( \beta \) to compute the total continuum light curves. This approach ensures that the wavelength-dependent diffuse continuum emission is properly accounted for alongside the accretion disc variability, providing a more complete picture of the observed continuum emission.
Figure~\ref{fig:figureSM5} show the effect of DCE on the observed variability compared to a pure accretion disc model.
The variability produced by the pure accretion disc model is consistent with the expected temporal broadening imposed by the disc transfer function alone. Incorporating a modest diffuse continuum emission (DCE) contribution of 3.9\% (as estimated from our photoionisation simulations) introduces a slight smoothing effect on the variability amplitude (magenta). In contrast, increasing the DCE fraction to 50\% (orange) or 100\% (red) results in excessive smoothing, which fails to reproduce the observed sharp features in the light curve.
The other photometric bands are not shown here, as even at 7705\text{\AA}, where the DCE contribution is 26.3\%, the resulting light curve would still not explain the sharper variability features observed.
These findings suggest that while a modest DCE component should be included for a more accurate description, it does not dominate the emission. The baseline accretion disc model (SS73 and lamp-post geometry) still provides a good qualitative match to the observed variability.

\end{appendix}

\end{document}